\documentclass[10pt,twoside]{classe-cmb}
\usepackage{amssymb,amsbsy,amsmath,amsfonts,amssymb,amscd}
\usepackage{latexsym,euscript,exscale,epic,eepic,epsfig}
\usepackage[english,francais]{babel}
\usepackage{times}
\usepackage[latin1]{inputenc}
\newcommand*{\beq}{\begin{equation}}
\newcommand*{\eeq}{\end{equation}}
\newcommand*{\beqa}{\begin{eqnarray}}
\newcommand*{\eeqa}{\end{eqnarray}}
\newcommand*{\ba}{\begin{array}}
\newcommand*{\ea}{\end{array}}

\newtheorem{e-proposition}[theorem]{Proposition}

\newtheorem{e-definition}[theorem]{Definition\rm}

\ComParit{S1296-2147}
\PIT{FLA}
\PXHY{????}
\Add{?}
\Volume{0}
\Year{2003}
\FirstPage{1}
\LastPage{??}
\AuteurCourant{}
\TitreCourant{ } 
\Journal
\Rubrique{Rubrique}{Heading}
\SousRubrique{Sous-rubrique}{Sub-Heading}
\PresentePar{First name}{NAME}
\Recu{blank}{}{}
\begin{document}
\bibliographystyle{unsrt}
\selectlanguage{english}
\TitleOfDossier{The Cosmic Microwave Background: 
\\present status and cosmological perspectives
}
\title{%
CMB Polarization as complementary information to anisotropies
}
\author{%
J. Kaplan~$^{\text{a}}$,\ \
J. Delabrouille~$^{\text{b}}$,\ \
P. Fosalba~$^{\text{c}}$,\ \
C. Rosset~$^{\text{d}}$
}
\address{%
\begin{itemize}\labelsep=2mm\leftskip=-5mm
\item[$^{\text{a}}$]
PCC Collège de France, Paris,
E-mail: kaplan@cdf.in2p3.fr
\item[$^{\text{b}}$]
PCC Collège de France, Paris,
E-mail: delabrou@cdf.in2p3.fr
\item[$^{\text{c}}$]
Institut for Astronomy/U Hawaii,
E-mail: fosalba@ifa.hawaii.edu
\item[$^{\text{d}}$]
PCC Collège de France, Paris,
E-mail: rosset@cdf.in2p3.fr
\end{itemize}
}
\maketitle
\thispagestyle{empty}
\begin{Abstract}{%
The origin of CMB polarization is reviewed. Special emphasis is
placed on the cosmological  information encoded in it: the nature of
primordial fluctuations, the connection with the inflation
paradigm. Insights into more recent epochs are also discussed: early
reionization and high redshift matter distribution from CMB
lensing.

{\bf\large Résumé.}  
L'origine de la polarisation du CMB est passée en revue. On insiste
plus particulièrement sur l'information qu'elle contient: la nature
des fluctuations primordiales, le lien avec le paradigme de
l'inflation. On discute aussi les aperçus apportés sur des époques plus
récentes~: la réionisation précoce et la distribution de matière pour des
hauts décalages vers le rouge.
}\end{Abstract} 
\par\medskip\centerline{\rule{2cm}{0.2mm}}\medskip
\setcounter{section}{0}
\selectlanguage{english}
\section{Introduction} The CMB fluctuations are polarized\cite{rees68a,BaskoPolnarev80}. The
information carried by CMB Polarization is one of the necessary inputs to
``precision cosmology''. In fact it puts strong additional constraints
on the cosmological parameters and provides crucial tests of the
relevance of inflation to the standard cosmological model.
In the following we describe the CMB polarization
 and the cosmological information it provides.

CMB polarization is extremely difficult to measure because it is small and
because the polarized foregrounds are poorly known. The experimental
situation is reviewed in another article in this issue \cite{delabrouille03}
\section{Polarization observables}
The polarization vector of an electromagnetic wave is described by the
electromagnetic field $\vec{E}$, orthogonal to the direction of
propagation $\vec{k}$. A general radiation is an incoherent
superposition of waves with the same wave vector $\vec{k}$.
Choosing two basis vectors $\vec{e}_1$ and $\vec{e}_2$ orthogonal to $\vec{k}$,
all statistical information is encoded in the the ``coherence matrix'' $C$:
\beq
C=\begin{pmatrix}
\langle |E_1|²\rangle &\langle E_1\,E_2^*\rangle \\ \langle E_2\,E_1^*\rangle &\langle |E_2|²\rangle 
\end{pmatrix} = \frac{1}{2}\begin{pmatrix} I+Q & U-i V \\ U+iV & I-Q
\end{pmatrix},
\eeq
where the Stokes parameters $I, Q, U, V$\footnote{For the coherence
  matrix and the Stokes parameters, see the classical textbooks \cite{chandrasekhar60,jackson99,born2000}.} satisfy the inequality $ I^2 \ge Q^2 + U^2 + V^2$, which means that
the polarized energy cannot exceed the total energy. It becomes an
equality for a  fully polarized radiation.

The $Q$ and $U$ Stokes parameters, which describe the linear polarization, depend on the reference frame. If
$\vec{e}_1$ and $\vec{e}_2$ are rotated by an angle $\theta$ around
$\vec{k}$ then $Q$ and $U$ rotate to $Q'$ and $U'$ by an angle
$2\theta$:
\beq
\begin{array}{l}
Q'=Q\,\cos 2\theta + U\,\sin 2\theta\\
U'=-Q\,\sin 2\theta + U\,\cos 2\theta 
\end{array}\qquad \mbox{ or } \qquad
Q\pm i\,U \rightarrow Q'\pm iU'=e^{\mp 2\,i\,\theta}\,(Q\pm i\,U).
\eeq
This is the transformation of a spin 2 object\footnote{For this
  section and the following, useful references are: \cite{kosowsky96}
  and \cite{ZaldarriagaSel97}}. The $V$ parameter describes the circular
polarization and is invariant under rotations.
\section{The origin of CMB polarization}
\begin{figure}[ht] {\centering
\epsfig{file=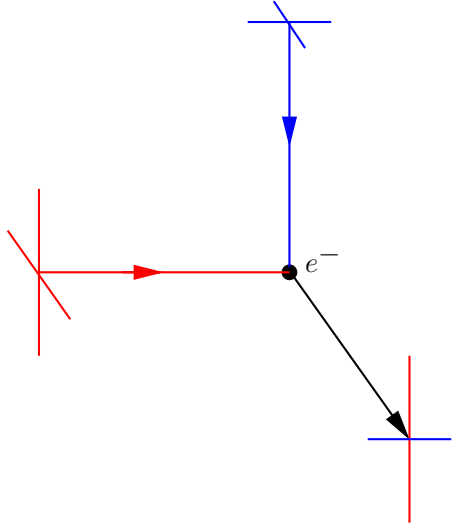,width=.4\textwidth}
\caption{Polarization from Thomson scattering\label{polgen}}}
\end{figure}

The CMB polarization originates from rescattering of the
primordial photons on the hot electron gas on their way to us.
A quadrupole anisotropy of the photon flux at one point on the last
scattering surface generates a polarization in the direction of
observation as shown in figure \ref{polgen}: the cross-section of
Thomson scattering is proportional to the square of the scalar product
$\epsilon1 . \epsilon2$ of the incoming and outgoing photons polarizations,
thus outgoing photons only carry polarization orthogonal to
the scattering plane.
Radiation fluxes from different directions are incoherent, therefore
intensities from 
opposite incident directions add up and only even multipoles
of the flux angular distribution contribute. In fact, only the monopole
and the quadrupole remain because the scattering cross-section
is quadratic in  $\epsilon1 . \epsilon2$. As $Q$ and $U$ are
intensity differences, they only get contribution from the
quadrupole. 

Note that any polarization gets averaged out by
successive rescatterings, therefore we only observe polarization from
the last one, at the end of recombination. The scale of the fluctuations
we can observe is of the order of the mean free path of photons when
the last scattering occurs and cannot be larger than the thickness of
the last scattering surface. 

\begin{figure}[ht]{\centering
\epsfig{file=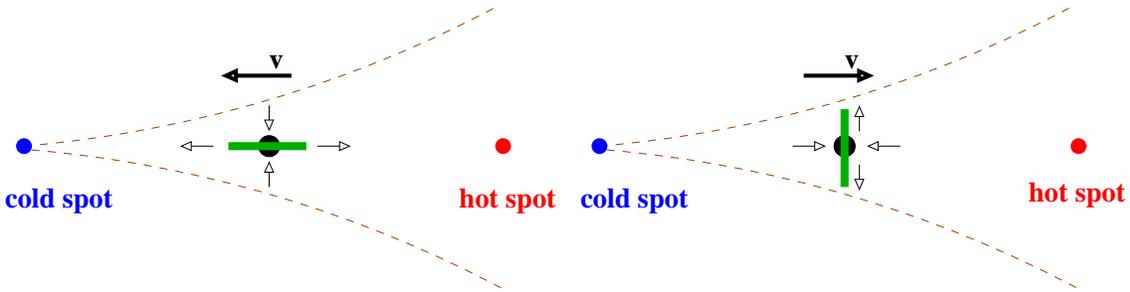,width=\textwidth}
\caption{Polarization directions from velocity gradients. Polarization
  directions when the fluid is accelerated towards the cold spot (left
  panel) or decelerated towards a hot spot (right panel). The brown
  dashed curves are fluid stream lines. The small thin arrows are the
  fluid velocities in the fluid rest frame near the scattering point.
The large thick arrows are the directions of the fluid motion near the
scattering point relative to the hot and cold spots.   \label{polv}}}
\end{figure}

For density fluctuations, the local quadrupole anisotropies of the
photon flux on the last scattering surface arise from velocity
gradients, as sketched in 
figure  \ref{polv}. In the photon baryon fluid rest frame, the
velocities of 
neighboring particles tend to diverge
radially from and converge transversely to 
the scattering point when the fluid is 
accelerated from a  hot spot (density dip, potential maximum) to a
cold spot (density peak, potential dip). The reverse velocity scheme
applies when the fluid is decelerated away from a cold spot. By 
Doppler shift, this induces a quadrupole flux anisotropy around the
last scattering point, leading to radial polarization in the first case
and to transverse polarization in the second case. This simple
geometrical scheme does not apply to primordial gravitational waves which are quadrupolar by nature,
and therefore do not need velocity gradients to generate CMB polarization.

The above mechanism does not produce any circular polarization, and
from now on, we forget about the $V$ Stokes parameter\footnote{The relativistic
  electron gas of foreground clusters may produce a very week circular
  polarization at very low frequencies \cite{cooray03}}.

\section{Polarized multipole expansion} 
As we are observing the celestial sphere, it is convenient to develop
the CMB fluctuation on spherical harmonics. Because $Q\pm i\,U$ are spin 2
objects, they must be developed on spin 2 spherical harmonics
\cite{Goldberg67}: 
\beq
(Q\pm i\,U)(\vec{n})=\sum_{l\ge 2,\ |m|\le l} a_{\pm 2 l m}\,_{\pm 2}Y_l^m(\vec{n})
\eeq

From these spin 2 objects, one can construct 2 real scalar quantities
with opposite behavior 
under parity transformations\footnote{In
  reference \cite{kosowsky96} $E$ ($B$) modes are noted $C$ ($G$) modes. For more geometric insights into the polarization patterns, see \cite{hu97b} and \cite{zaldar2001EB} }:
\beq
\begin{array}{l}
E(\vec{n})=\sum_{l\ge 2,\ |m|\le l} a^E_{l\,m}\,Y_l^m(\vec{n})\qquad \mbox{ with positive parity}\\
B(\vec{n})=\sum_{l\ge 2,\ |m|\le l} a^B_{l\,m}\,Y_l^m(\vec{n})\qquad \mbox{ with negative parity},
\end{array}
\ \ \mbox{ where } \  \ 
\begin{array}{l}
a^E_{l\,m}=-\frac{a_{2 l m}+a_{-2 l m}}{2}\\
a^B_{l\,m}=i\frac{a_{2 l m}-a_{-2 l m}}{2}.
\end{array}
\eeq
\begin{figure}[h]{\centering
\epsfig{file=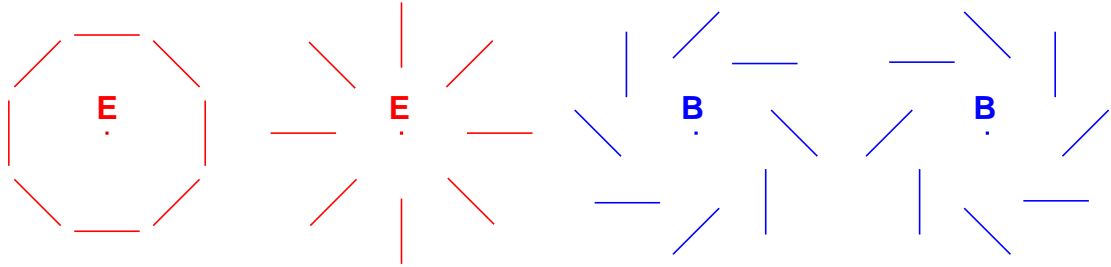,width=0.98\textwidth,clip}
\caption{Typical $E$ or $B$ type polarization patterns. The opposite
  transformation   properties under reflection 
  appear clearly (adapted from Zaldarriaga \cite{Zaldarriaga03PolCMB}).  
 \label{polpattern}}}
\end{figure} 
The different parity properties of $E$ and $B$ type polarization
fluctuation are illustrated in figure \ref{polpattern}. They 
arise from different parity properties of the velocity gradients, and
  trace the geometrical properties of the underlying fluctuations. 
 
\begin{figure}[h]
\begin{center}
\epsfig{file=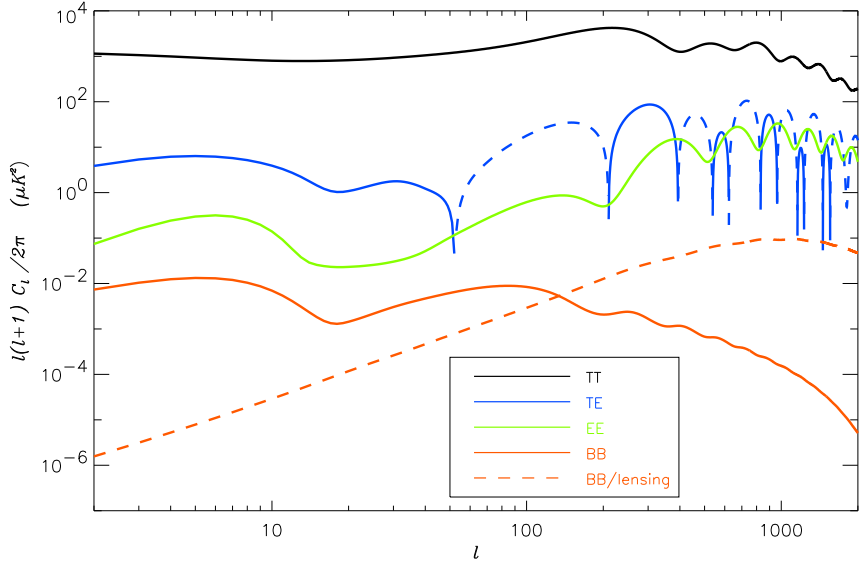,width=.8\textwidth}
\caption{The four observable spectra $TT,\ TE,\ EE,\
  BB$. The $TE$ spectrum can be negative (dotted segments), and 
  apparent cusps occur when $TE$ changes sign. They correspond to
  the compression and expansion extrema of the $TT$ spectrum, when the
  fluid velocity vanishes. Note that the $EE$ spectrum is in phase
  opposition relative to $TT$. These spectra have been computed using
  the code CMBFAST \cite{seljak96} (http://www.cmbfast.org). 
}\label{polspec}
\end{center}
\end{figure}

Including polarization, the statistical properties of Gaussian CMB
fluctuations are characterized by four power spectra:
\beq
C^{TT}_l=\langle {a^T_{lm}}^*a^T_{lm} \rangle,\qquad
C^{ET}_l=\langle {a^E_{lm}}^*a^T_{lm} \rangle,\qquad
C^{EE}_l=\langle {a^E_{lm}}^*a^E_{lm} \rangle,\qquad
C^{BB}_l=\langle {a^B_{lm}}^*a^B_{lm} \rangle
\eeq
An example of the four expected spectra as predicted by the best model
compatible with the results of the WMAP experiment
\cite{wmap2003basic} is displayed on
figure \ref{polspec}.
Notice that, because of the negative parity of $B$, its cross power spectra
with both $T$ and $E$ are expected to vanish.

\section{Information carried by CMB polarization}
The CMB temperature fluctuations are the imprint on the last
scattering surface of density and metric perturbations. They are
classified as ``scalar", ``vector'' and ``tensor'' depending on their
transformation properties under rotations.  Vector perturbation get
damped by expansion 
and only the scalar and tensor ones survive.
The scalar contributions comprise total density fluctuations, also
called curvature or adiabatic perturbations, and isocurvature
fluctuations where only the relative densities of the different
components vary. Tensor fluctuations are expected from the primordial
gravity waves predicted in the inflationary scenario. CMB polarization
allows to separate them  from scalar perturbations (see e.g. \cite{Hu03cmb,Zaldarriaga03PolCMB}).

\begin{figure}{\centering
  \epsfig{file=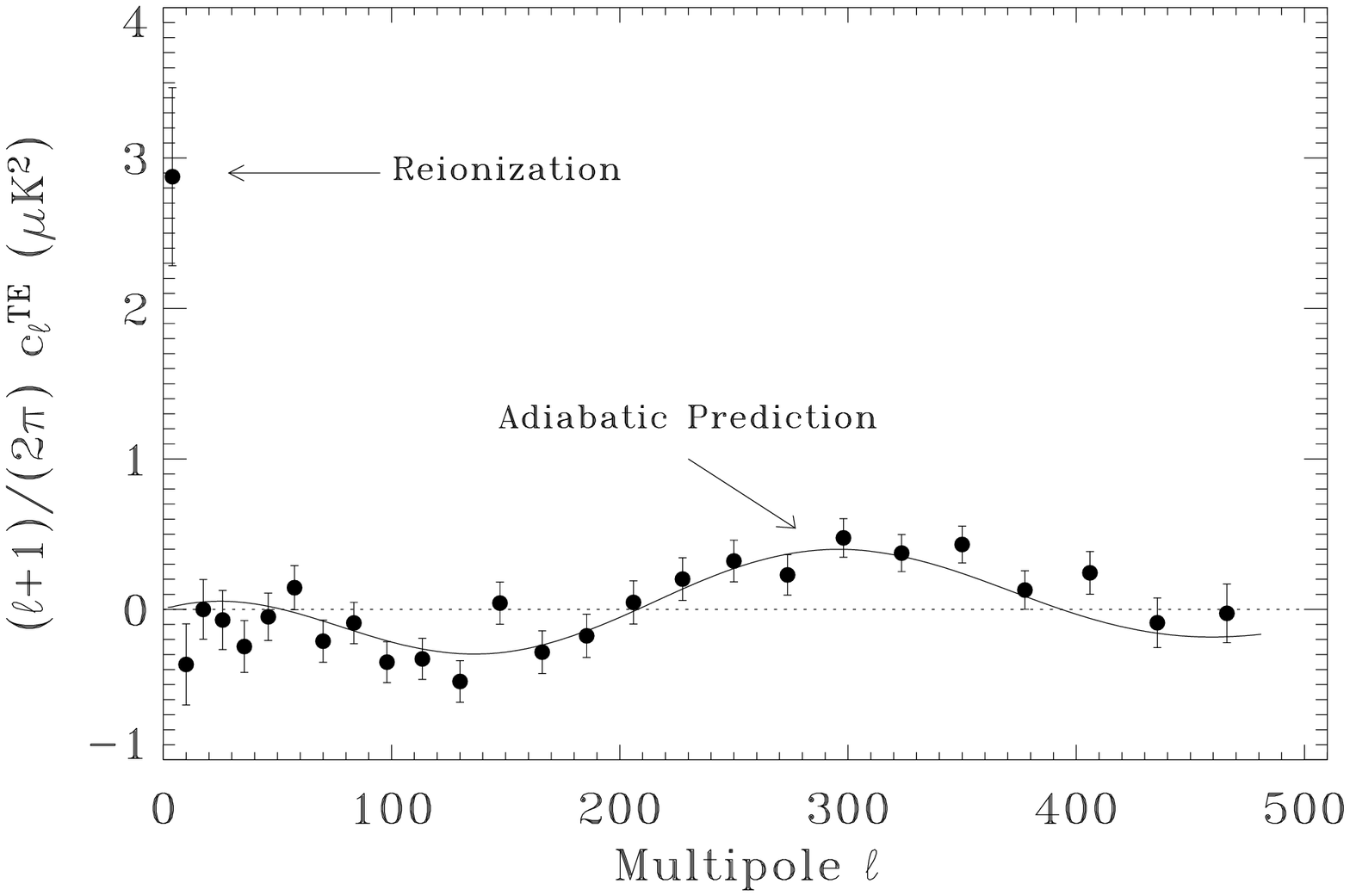,width=.7\textwidth}
\caption{The $TE$ spectrum as observed by WMAP. The 
  solid  line is
  computed from the best fit to the $TT$ spectrum, without additional free
  parameter, assuming adiabatic perturbations (Figure from reference
  \cite{wmap2003te}, reproduced with permission of the AAS).}\label{polwmap}} 
\end{figure}

{\bf E polarization: \bf  scalar fluctuations.}
Scalar perturbations  only generate positive parity polarization
patterns, and therefore can only produce $E$ type CMB polarization
fluctuations. In the absence of isocurvature perturbations, the $TE$
spectrum is completely calculable from the $EE$ spectrum. This
property is extremely well verified by the $TE$ spectrum  observed by
the WMAP collaboration (Figure \ref{polwmap}).  

{\bf B Polarization: tensor fluctuations.} On the contrary, tensor
fluctuation can produce both $E$ 
and $B$ polarization \cite{SelZar97GRW,KamKos98GRW}. Therefore the
relative intensity of B polarization modes 
is a constraint  on the ratio $r$ of tensor to scalar fluctuations.

\subsection{Link with the inflation paradigm }
The inflation paradigm\footnote{For a thorough
    introduction to inflation, see e.g. \cite{LiddleLyth00Cosmo}} was
  introduced by Guth \cite{Guth81Infl}, Linde 
\cite{linde82infl} and Albrecht \& Steinhardt \cite{AlbrSteinh82Infl}
to solve the horizon problem: 
the present observed Universe appears homogeneous on scales which have
never been in causal contact, due to the finite light speed. In
particular the CMB is homogeneous across the whole sky at the
$10^{-5}$ level, although the ``horizon'' size at the time of
decoupling is of order 1 degree on the sky sphere. 

In inflationary models, the very early Universe undergoes a rapid
expansion phase driven by the vacuum energy of some scalar
field. This inflation must last long enough for the comoving size of
the horizon before inflation to be larger than the present size of the
horizon ($\simeq 1/H_0$). This implies an inflation factor of order
$e^{40}$ to $e^{60}$, depending on when inflation ends (after GUT symmetry
breaking and before baryogenesis). 
 
Although the horizon problem, the near flatness of our universe and
the elimination of unwanted relics (GUT's monopoles, domain walls ...)
were the original motivations to 
propose inflation, this scenario also naturally provides the seeds of the
present large scale structures.
These seeds are the microscopic quantum fluctuations of the inflaton
field, which are inflated to macroscopic adiabatic scalar perturbations
during inflation. One of the generic predictions of inflation is the
presence of acoustic peaks in the $C_l$ spectrum which have now been
observed beyond any doubt
\cite{Saskatoon97,Maxima1_00_ab,Boomerang1_00_ab,Archeops_CL_ab}. The
reason is that  
perturbations begin to oscillate when they enter the horizon and all
perturbations with the same size do so at the same time. The acoustic
peaks arise from this coherent oscillation. The fact that the peaks in
the $TE$ spectrum observed by WMAP are out of phase with the
temperature peaks is a further confirmation of the acoustic scenario
because polarization originates in  velocity gradients. The dip in the
$TE$ spectrum around $l\simeq 150$ indicates the presence of
superhorizon size fluctuations at the time of decoupling, which is a
strong indication in favor of inflation.     

In much the same way, the space metric also undergoes  quantum
fluctuations which produce primordial gravity waves (tensor
perturbations). The ratio between tensor and scalar
perturbations, defined as  $r=\left.\frac{C_l^T}{C_l^S}\right|_{l=2}$, is
related to the energy scale of inflation $E_\mathrm{inflation}$. For
example in the popular ``slow-roll'' approximation (see
e.g. \cite{LiddleLyth00Cosmo}),
$$
r=0.1\,\left(\frac{E_\mathrm{inflation}}{2.\,10^{16}\, GeV}\right)^4
$$
Therefore, detection of
$B$-mode polarization will allow to measure this energy scale, but the
quick decrease of $r$ with $E_\mathrm{inflation}$ may make it difficult.
The primordial $BB$ spectrum in figure \ref{polspec} (solid red line)
corresponds to $r=0.1$. 
\section{Distortions by foregrounds}
The polarized CMB spectra will be affected by various foregrounds.
Apart from those related to galactic dust an synchrotron
emission, which we shall not discuss here, some foreground effects,
such as reionization and gravitational lensing, have an important
cosmological significance. 

\subsection{Reionization}
\begin{figure}[th]{\centering
\epsfig{file=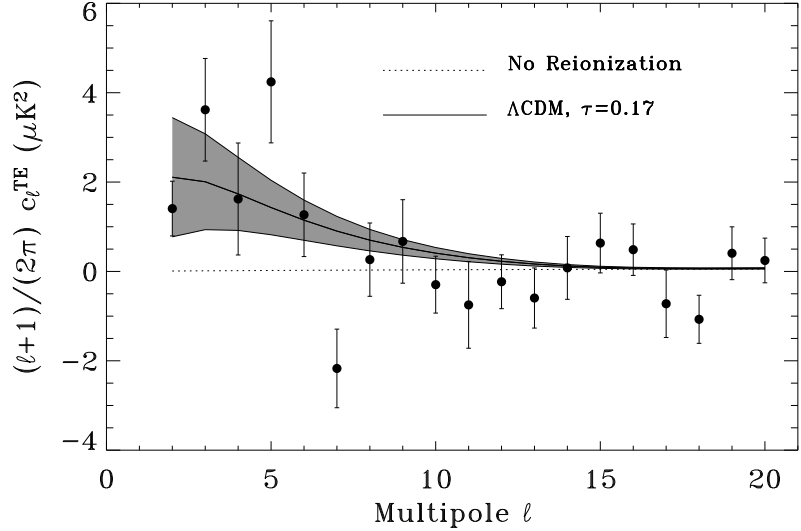,width=.8\textwidth}
\caption{Zoom on the low $l$ region of the $TE$ spectrum as observed
  by WMAP. Dotted line:  $\Lambda$CDM model
  with no reionization. Solid line: $\Lambda$CDM plus a reionization
  optical depth $\tau = 0.17$. Grey area: the one sigma
  uncertainty from cosmic variance (Figure from reference \cite{wmap2003te}, reproduced with permission of the AAS).}\label{polwmap2}}
\end{figure}
It has been known for a long time \cite{GunnPet65Reio} that the universe
is observed to be ionized, at least up to a redshift of $z\simeq
6$ (see e.g. \cite{LoebBark01Reio}). This
reionization is expected from the high energy photons produced in the
collapse of the first generation of stars (for recent reviews, see
\cite{barkanaLoeb01}) and \cite{miralda03}). Present
estimations place the beginning of reionization somewhere 
between $z=7$ and $z=30$ \cite{Hu03cmb}.   Reionization distorts the
spectrum of CMB perturbations: free electrons  re-scatter the CMB
photons, diluting somewhat the temperature
fluctuations. Exactly as on the last scattering surface, the local CMB
quadrupole causes the rescattered CMB photons to be polarized.
Because photons stream freely between their last scattering
at recombination and their rescattering 
on reionized electrons, the polarized CMB fluctuations from
reionization appear at larger scale, and therefore induce a low $l$ peak in the polarized power spectra. This reionization bump, below $l=20$, carries
rich information on the reionization history of the universe
\cite{zaldarriaga97,kapling03,holder03}. The low $l$ peak appears
 on the $TE$, $EE$, $BB$ spectra in figure \ref{polspec},
where the reionization optical depth has been assumed to be 0.17
\cite{wmap2003te}. 
The WMAP experiment \cite{wmap2003basic,wmap2003te} has claimed a first
detection of polarization from reionization in the $TE$ spectrum
(figure \ref{polwmap2}).
They evaluate the reionization optical depth to be $\tau=0.17\pm
0.04$, larger than expected. This would be well described by an early
reionization at redshift $11<z_r<30$ at the 95\% confidence
level. These results have trigered much work on early
reionization, see for instance
\cite{HuHolder03,barkana03,knox03,PengHai03,onken03}.
Forthcoming results of WMAP on the $EE$ spectrum will allow more
precise statements. Future high precision measurements of large scale
CMB polarization will be a very useful tool to understand early
star formation. 
\subsection{Lensing of primordial polarized fluctuations: small scale
  B modes}
One of the main distortions of the polarized CMB
fluctuations spectrum arises from the lensing of the CMB photons by matter
inhomogeneities on their way to us \cite{Seljak96Lensing,Bernardeau97WeakL}. Lensing does not produce
nor rotate polarization, but deflection changes its spatial
patterns. In particular it blurs the parity properties, thereby
leaking $E$ modes into $B$ modes \cite{ZalSel98LensPol}. In figure \ref{polspec}, the dashed red
line represents the $BB$ spectrum expected from lensing. One can see
that it  dominates at small scale and will do so at large scale if the
energy scale of inflation is too low. Knox and Song \cite{KnoxSong02},
and independently Kesden, Cooray and Kamionkowski \cite{KesCooKam02GWSep} 
have recently shown that $B$ modes from primordial gravity waves will
be masked by lensing  $B$ modes if the energy scale of inflation lies
below $3.2\, 10^{15}$ GeV\footnote{For a recent update on the
detectability of $B$ modes in inflationary models, see \cite{kinney03}}.

Lensing of the CMB should not be only considered as a nuisance to
be subtracted. It can be used as a tool to study  the mass
distribution between us and the last scattering surface. 
 Hu and Okamoto \cite{HuOka01,OkaHu03} and also Kesden, Cooray, and
 Kamionkowski \cite{KesCooKam03LensRec} have devised an approach which
 treats the 
CMB statistically, but considers the foreground lensing potential as
fixed. The breaking of anisotropy allows to consider $TB$ and $EB$
correlations which, combined with the the usual ones, can be used to
reconstruct the lensing potential. This may provide a powerful tool
to investigate the high redshift mass distribution.

\section{Conclusion}
The flatness of the universe and the harmonic structure of the CMB
spectrum are now well established. The first observations of CMB
polarization by DASI \cite{kovac02:dasi} and WMAP \cite{wmap2003te}
  point in the direction of inflation. We have seen that CMB
  polarization observations have a high potential to assess the
  cosmological scenario and in particular have much to say about
  inflation. Moreover, it carries information on early star formation
  and high redshift matter distribution. New results on all these
  issues will soon come from observations in progress 
  or in preparation, see \cite{delabrouille03} in this issue.

\Acknowledgements{We would like to thank François-Xavier Desert for many useful suggestions.}
%
\bibliography{mnemokap,cmb}

\begin{thebibliography}{10}

\bibitem{rees68a}
M.J. Rees.
\newblock Polarization and spectrum of the primeval radiation in an anisotropic
  universe.
\newblock {\em ApJ}, 153:L1, 1968.

\bibitem{BaskoPolnarev80}
M.~M. {Basko} and A.~G. {Polnarev}.
\newblock {Polarization and anisotropy of the RELICT radiation in an
  anisotropic universe}.
\newblock {\em MNRAS}, 191:207--215, April 1980.

\bibitem{delabrouille03}
J.~Delabrouille, J.~Kaplan, M.~Piat, and C.~Rosset.
\newblock Polarization experiments.
\newblock In {\em The Cosmic Microwave Background: present status and
  cosmological perspectives}, This issue of C. R. (Physique). Academie des
  Sciences, Paris, 2003.

\bibitem{chandrasekhar60}
S.~Chandrasekhar.
\newblock {\em {Radiative Transfer}}.
\newblock Dover Publications, 1960.

\bibitem{jackson99}
J.~D. Jackson.
\newblock {\em Classical Electrodynamics}.
\newblock John Wiley \& Sons, 3rd edition, 1999.

\bibitem{born2000}
M.~Born and E.~Wolf.
\newblock {\em Principles of Optics}.
\newblock Pergamon Press, 7th edition, 2000.

\bibitem{kosowsky96}
A~Kosowsky.
\newblock {Cosmic Microwave Background Polarization}.
\newblock {\em Annals Phys.}, 246:49, 1996.

\bibitem{ZaldarriagaSel97}
M.~Zaldarriaga and U.~Seljak.
\newblock {An All-Sky Analysis of Polarization in the Microwave Background}.
\newblock {\em Phys. Rev. D}, 55:1830, 1997.

\bibitem{cooray03}
A.~Cooray, A.~Melchiorri, and J.~Silk.
\newblock {Is the Cosmic Microwave Background Circularly Polarized?}
\newblock {\em Phys.Lett. B}, 554:1, 2003.

\bibitem{Goldberg67}
J.N. Goldberg et~al.
\newblock {Spin-s Spherical Harmonics and\ \ \ \'\!\!\!\!$\partial$}.
\newblock {\em Jour. Math. Phys.}, 8:2155, 1967.

\bibitem{hu97b}
W.~Hu and M.~White.
\newblock {A CMB Polarization Primer}.
\newblock {\em New Astronomy}, 2:323, 1997.

\bibitem{zaldar2001EB}
M.~Zaldarriaga.
\newblock {The nature of the E-B decomposition of CMB polarization}.
\newblock {\em Phys.Rev.}, D64:103001, 2001.

\bibitem{Zaldarriaga03PolCMB}
M.~Zaldarriaga.
\newblock {\em {The Polarization of the Cosmic Microwave Background}}, volume
  Vol 2: Measuring and modelling the universe of {\em Carnegie Observatories
  Astrophysics Series}.
\newblock Cambridge University Press, 2003.
\newblock astro-ph/0305272.

\bibitem{wmap2003basic}
{C.L.} Bennett et~al.
\newblock {First Year Wilkinson Microwave Anisotropy Probe (WMAP) Observations:
  Preliminary Maps and Basic Results}.
\newblock {\em ApJS}, 148, 2003.

\bibitem{seljak96}
U.~Seljak and M.~Zaldarriaga.
\newblock {A line of sight approach to Cosmic Microwave Background
  anisotropies}.
\newblock {\em ApJ}, 469:437, 1996.

\bibitem{Hu03cmb}
W.~Hu.
\newblock {CMB Temperature and Polarization Anisotropy Fundamentals}.
\newblock {\em Annals Phys.}, 303:203, 2003.

\bibitem{wmap2003te}
A.~Kogut et~al.
\newblock {Wilkinson Microwave Anisotropy Probe (WMAP) First Year Observations:
  TE Polarization}.
\newblock {\em ApJS}, 148, 2003.

\bibitem{SelZar97GRW}
U.~Seljak and M.~Zaldarriaga.
\newblock {Signature of Gravity Waves in Polarization of the Microwave
  Background}.
\newblock {\em Phys.Rev.Lett.}, 78:2054, 1997.

\bibitem{KamKos98GRW}
M.~Kamionkowski and A.~Kosowsky.
\newblock {Detectability of Inflationary Gravitational Waves with Microwave
  Background Polarization}.
\newblock {\em Phys.Rev.}, D57:685, 1998.

\bibitem{LiddleLyth00Cosmo}
A.~R. Liddle and D.~H. Lyth.
\newblock {\em {Cosmological inflation and large-scale structure}}.
\newblock Cambridge University Press, 2000.

\bibitem{Guth81Infl}
{A. H.} Guth.
\newblock {Inflationary universe: A possible solution to the horizon and
  flatness problems}.
\newblock {\em Phys. Rev. D}, 23:347, 1981.

\bibitem{linde82infl}
A.D. Linde.
\newblock {A new inflationary universe scenario: a possible solution of the
  horizon, flatness, homogeneity, isotropy, and primordial monopole problems}.
\newblock {\em Phys. Lett.}, 108B:389, 1982.

\bibitem{AlbrSteinh82Infl}
A.~Albrecht and {P. J.} Steinhardt.
\newblock {Cosmologie for grand unified theories with radiation induced
  symmetry breaking}.
\newblock {\em Phys. Rev. Lett.}, 48(1220), 1982.

\bibitem{Saskatoon97}
C.~B. {Netterfield}, M.~J. {Devlin}, N.~{Jarolik}, L.~{Page}, and E.~J.
  {Wollack}.
\newblock {A Measurement of the Angular Power Spectrum of the Anisotropy in the
  Cosmic Microwave Background}.
\newblock {\em ApJ}, 474:47, January 1997.

\bibitem{Maxima1_00_ab}
S.~{Hanany} et~al.
\newblock {MAXIMA-1: A Measurement of the Cosmic Microwave Background
  Anisotropy on Angular Scales of 10'-5{$\deg$}}.
\newblock {\em ApJL}, 545:L5, December 2000.

\bibitem{Boomerang1_00_ab}
P.~{de Bernardis} et~al.
\newblock {A flat Universe from high-resolution maps of the cosmic microwave
  background radiation}.
\newblock {\em Nature}, 404:955--959, April 2000.

\bibitem{Archeops_CL_ab}
A.~Beno{\^ i}t et~al.
\newblock {The Cosmic Microwave Background Anisotropy Power Spectrum measured
  by Archeops}.
\newblock {\em A\&A}, 399:L19, 2003.

\bibitem{GunnPet65Reio}
J.~E. Gunn and B.~A. Peterson.
\newblock {On the Density of Neutral Hydrogen in Intergalactic Space}.
\newblock {\em ApJ}, 142:1633, 1965.

\bibitem{LoebBark01Reio}
A.~Loeb and R.~Barkana.
\newblock {The Reionization of the Universe by the First Stars and Quasars}.
\newblock {\em ARA\&A}, 39:19, 2001.

\bibitem{barkanaLoeb01}
Barkana and A.~Loeb.
\newblock {In the Beginning: The First Sources of Light and the Reionization of
  the Universe}.
\newblock {\em Phys. Rep.}, 349:125, 2001.

\bibitem{miralda03}
J.~Miralda-Escude.
\newblock {The Dark Age of the Universe}.
\newblock {\em Science}, 300:1904, 2003.

\bibitem{zaldarriaga97}
M.~Zaldarriaga.
\newblock {Polarization of the Microwave Background in reionized models}.
\newblock {\em Phys.Rev.D}, 55:1822, 1997.

\bibitem{kapling03}
M.~{Kaplinghat}, M.~{Chu}, Z.~{Haiman}, G.~P. {Holder}, L.~{Knox}, and
  C.~{Skordis}.
\newblock {Probing the Reionization History of the Universe using the Cosmic
  Microwave Background Polarization}.
\newblock {\em ApJ}, 583:24--32, January 2003.

\bibitem{holder03}
G.~Holder, Z.~Haiman, M.~Kaplinghat, and L.~Knox.
\newblock {The Reionization History at High Redshifts II: Estimating the
  Optical Depth to Thomson Scattering from CMB Polarization}.
\newblock {\em to appear in ApJ}, 2003.

\bibitem{HuHolder03}
W.~Hu and G.P. Holder.
\newblock {Model-Independent Reionization Observables in the CMB}.
\newblock {\em Phys.Rev.}, D68:023001, 2003.

\bibitem{barkana03}
R.~Barkana and A.~Loeb.
\newblock {GRBs versus Quasars: Lyman-alpha Signatures of Reionization versus
  Cosmological Infall}.
\newblock {\em submitted to ApJ}, 2003.
\newblock astro-ph/0305470.

\bibitem{knox03}
L.~Knox.
\newblock {CMB Signatures of Extended Reionization}.
\newblock In S.~Hanany and K.A. Olive, editors, {\em "The Cosmic Microwave
  Background and its Polarization"}, To appear in New Astronomy Reviews, 2003.
\newblock astro-ph/0305588.

\bibitem{PengHai03}
S.~{Peng Oh} and Z.~Haiman.
\newblock {Fossil HII Regions: Self-Limiting Star Formation at High Redshift}.
\newblock {\em Submitted to MNRAS}, 2003.
\newblock astro-ph/0307135.

\bibitem{onken03}
Ch.~A. Onken and Jordi Miralda-Escude.
\newblock {History of Hydrogen Reionization in the Cold Dark Matter Model}.
\newblock {\em submitted to ApJ}, 2003.
\newblock astro-ph/0307184.

\bibitem{Seljak96Lensing}
U.~Seljak.
\newblock {Gravitational Lensing Effect on Cosmic Microwave Background
  Anisotropies: A Power Spectrum Approach}.
\newblock {\em ApJ}, 463(1), 1996.

\bibitem{Bernardeau97WeakL}
F.~Bernardeau.
\newblock {Weak lensing detection in CMB maps}.
\newblock {\em A\&A}, 324:15, 1997.

\bibitem{ZalSel98LensPol}
M.~Zaldarriaga and U.~Seljak.
\newblock {Gravitational lensing effect on Cosmic Microwave Background
  polarization}.
\newblock {\em Phys. Rev. D}, 58:23003, 1998.

\bibitem{KnoxSong02}
L.~Knox and Y.-S. Song.
\newblock {A limit on the detectability of the energy scale of inflation}.
\newblock {\em Phys.Rev.Lett.}, 89:011303, 2002.
\newblock astro-ph/0202286.

\bibitem{KesCooKam02GWSep}
M.~{Kesden}, A.~{Cooray}, and M.~{Kamionkowski}.
\newblock {Separation of Gravitational-Wave and Cosmic-Shear Contributions to
  Cosmic Microwave Background Polarization}.
\newblock {\em Physical Review Letters}, 89(1):011304, July 2002.

\bibitem{kinney03}
W.H. Kinney.
\newblock {The energy scale of inflation: is the hunt for the primordial B-mode
  a waste of time?}
\newblock In S.~Hanany and K.A. Olive, editors, {\em The Cosmic Microwave
  Background and its Polarization}, To appear in New Astronomy Reviews, 2003.
\newblock astro-ph/0307005.

\bibitem{HuOka01}
W.~Hu and T.~Okamoto.
\newblock {Mass Reconstruction with CMB Polarization}.
\newblock {\em ApJ}, 574:566, 2002.

\bibitem{OkaHu03}
T.~Okamoto and W.~Hu.
\newblock {CMB Lensing Reconstruction on the Full Sky}.
\newblock {\em Phys. Rev.}, D67:083002, 2003.

\bibitem{KesCooKam03LensRec}
M.~Kesden, A.~Cooray, and M.~Kamionkowski.
\newblock {Lensing Reconstruction with CMB Temperature and Polarization}.
\newblock {\em Phys.Rev.}, D67:123507, 2003.

\bibitem{kovac02:dasi}
J.~M. Kovac, E.~M. Leitch, C.~Pryke, J.~E. Carlstrom, N.~W. Halverson, and
  W.~L. Holzapfel.
\newblock {Detection of polarization in the cosmic microwave background using
  DASI}.
\newblock {\em Nature}, 420:772, 2002.

\end{thebibliography}
\end{document}